# How Different Is It Between Machine-Generated and Developer-Provided Patches?

An Empirical Study on The Correct Patches Generated by Automated Program Repair Techniques

Shangwen Wang[1,3], Ming Wen[2], Liqian Chen[1,3], Xin Yi[1,3], and Xiaoguang Mao[1,3]
[1]National University of Defense Technology, Changsha, China
[2]The Hong Kong University of Science and Technology, Hong Kong, China
[3]Hunan Key Laboratory of Software Engineering for Complex Systems, Changsha, China
{wangshangwen13, lqchen, yixin09, xgmao}@nudt.edt.cn; mwenaa@cse.ust.hk

*Abstract*—<u>Background</u>: Over the years, Automated Program Repair (APR) has attracted much attention from both academia and industry since it can reduce the costs in fixing bugs. However, how to assess the patch correctness remains to be an open challenge. Two widely adopted ways to approach this challenge, including manually checking and validating using automated generated tests, are biased (i.e., suffering from subjectivity and low precision respectively). <u>Aim</u>: To address this concern, we propose to conduct an empirical study towards understanding the correct patches that are generated by existing state-of-the-art APR techniques, aiming at providing guidelines for future assessment of patches. <u>Method</u>: To this end, we first present a Literature Review (LR) on the reported correct patches generated by recent techniques on the Defects4J benchmark and collect 177 correct patches after a process of sanity check. We investigate how these machine-generated correct patches achieve semantic equivalence, but syntactic difference compared with developer-provided ones, how these patches distribute in different projects and APR techniques, and how the characteristics of a bug affect the patches generated for it. <u>Results</u>: Our main findings include 1) we do not need to fix bugs exactly like how developers do since we observe that 25.4% (45/177) of the correct patches generated by APR techniques are syntactically different from developer-provided ones; 2) the distribution of machine-generated correct patches diverges for the aspects of Defects4J projects and APR techniques; and 3) APR techniques tend to generate patches that are different from those by developers for bugs with large patch sizes. <u>Conclusion</u>: Our study not only verifies the conclusions from previous studies but also highlights implications for future study towards assessing patch correctness.

## I. INTRODUCTION

Automated Program Repair (APR) techniques, which are proposed to reduce the onerous burden of debugging [1] and increase software quality, are of tremendous value. APR techniques can be generally divided into two families including search-based [2-5] (also known as generate-and-validate) and synthesize-based [6-9] approaches, classified by how they generate candidate patches and traverse the search space.

The basic pipeline that most of APR systems follow contains three steps: fault localization, patch generation, and patch validation. In the first step, the APR system identifies suspicious code entities in a given program as the potential fault locations. Usually, it receives a list of statements ranked by the suspicious values calculated by Fault Localization (FL) techniques [10-13]. In the second step, given a fault location, the APR system tries to generate candidate patches by modifying the program. Then, in the last step, the APR system assesses whether the generated patch is correct, i.e., whether it fixes the defect. If the patch does not pass the validation, the second and third steps will be repeated until a valid patch is found or a predefined limitation is reached, e.g., the execution time. Over the years, many studies have been conducted with the aim to better identify the fault location [10-19], advance the patch generation process [2-9, 20-29], and enhance the assessment of patch correctness [30-34]. The scope of this paper belongs to the last one.

Traditionally, test cases are used as the criteria for judging if a generated patch is correct: a patch is considered as correct if it passes all the test cases [2, 3, 35]. However, this method is biased and inefficient as pointed out by the study [36] that the test suites in real world systems are usually weak such that most of the patches that pass all tests are incorrect. This problem, which is often referred to as patch overfitting [31, 32, 34], motivates the need of new methodologies for patch correctness assessment. To address this concern, recent works mainly follow two methods for evaluating patch correctness. One is utilizing an independent test suite generated by automatic test generation tools to verify the patch correctness [24, 31, 34, 37]. Following this method, a patch is labeled as correct if it passes both the original associated test suites and the newly generated one. However, a recent study has shown that independent test suite is not suitable for being used to evaluate the patch correctness alone [30] since it achieves low precision in discerning incorrect patches. The other method is author annotation, i.e., authors of APR techniques manually check the correctness of patches generated by their own tools [20-23, 25-28]. Following this method, a patch is labeled as correct if the authors consider it semantic equivalent to the developer-provided patch. Although this method achieves high effectiveness [30], it still faces the challenge of being subjective [21, 30] (also known as author bias).

In this paper, we conduct an empirical study on the correct patches that have already been generated by the state-of-the-art APR techniques, aiming at providing guidelines for author annotation in the future to reduce the bias of this process. We collect totally 177 patches generated by 10 state-of-the-art APR techniques evaluated on the benchmark Defects4J [38], all of which have been labeled as correct by both the APR tools' authors and our sanity check via manual investigation. Specifically, we seek to answer the following three research questions in this study:

**RQ1** *How do machine-generated correct patches differ from developer-provided ones?*

**RQ2** *How do different types of patches distribute?*

**RQ3** *Do APR tools tend to generate correct patches but different from the developer-provided ones for bugs with certain characteristics?*

A patch is generated based on the buggy location identified by fault localization techniques (i.e., denoted as edit point in this study) with certain code modifications. Based on this, the differences between patches can be distinguished in terms of two aspects, edits points and code modifications. To answer RQ1, we compare the collected patches with developer-provided ones and classify them into four types based on the aforementioned two aspects. We further investigate how the patches that are syntactically different from developer-provided ones achieve semantic equivalence. In RQ2, we investigate the distribution of patches from two aspects (i.e., different Defects4J projects and APR techniques) and observe that fault localization is critical for generating correct patches for bugs in three projects of Defects4J. In RQ3, we aim at investigating whether correlations exist between bug characteristics and the likelihood of APR tools to generate different patches from the developer-provided ones. By further analyzing the results and answers to these research questions, we distill several implications for future study towards the assessment of patch correctness, e.g., synthesis-based techniques such as Nopol are prone to generate different patches from the ground truth (i.e., the developer-provided patch).

We summarize our contributions in this study as below:
- We are the first to systematically study the correct patches that have been generated by APR techniques. We investigate how the patches that are different from the developer-provided ones achieve syntactic difference but semantic equivalence.
- We investigate the distribution of these patches from two aspects (i.e., Defects4J projects and APR techniques) and observe that the distribution of correct patches diverges for both of the two aspects.
- We study the correlation between the characteristics of a bug and the patches generated for the bug. The results reveal that correct patches different from the ground truth are more likely to be generated for bugs with large patch sizes.

The rest of the paper is organized as follows. Section II presents the background on various APR techniques and existing methods used for patch correctness assessment. We describe our study design in Section III. The results and analyses are presented in Section IV. Section V discusses implications of our findings and threats to validity. Section VI introduces related work. We conclude and describe future work in Section VII.

## II. BACKGROUND

In this section, we first present background information about automated program repair (APR) techniques. We subsequently elaborate methods that have been devised for assessing patch correctness in APR researches.

### A. APR Techniques

GenProg [2] is one of the first APR techniques that sparks the interests in APR. Given a buggy program and a set of tests, at least one of which is failing, it generates a population of repair candidates by using a number of mutation operators, such as statement deletion, insertion, and replacement. It then uses genetic programming to evolve the buggy program until a candidate program passing all the tests is found or a predefined time budget is reached. RSRepair [3] uses random search instead of genetic programming to traverse the search space of candidate solutions. It limits its patches to a single edit. Experimental results show that RSRepair is more efficient than GenProg in terms of time and test case evaluations [3]. AE [39] introduces a deterministic repair algorithm based on the insights that tests and candidates can be selected based on execution histories. This algorithm reduces the search space by an order of magnitude compared with GenProg. Kali [32] is a naive APR technique, which only deletes functionality. Although being simple, this technique has been shown to be as effective and efficient as GenProg, RSRepair, and AE [32].

The aforementioned techniques are all designed for C language. Recently, substantial APR techniques are designed for Java language. PAR [40] is a prominent APR technique which is based on a set of predefined human-provided patch templates. This technique has been shown to be able to fix the majority of its benchmark defects with only two templates (i.e., *Null Pointer Checker* and *Condition Expression Adder/Remover/Replacer*) [41]. Nopol [42] is an automatic repair tool focusing on branch conditions. It identifies branch statement directions that can pass negative test cases and then uses Satisfiability Modulo Theory (SMT) solvers to generate patches for the branch condition. ACS [20] also focuses on synthesizing patches for buggy if-conditions. Unlike Nopol, ACS attempts to rank the fix candidates using various ranking heuristics. JFix [9] adopts symbolic execution to infer specifications serving for patch synthesis. SimFix [28] takes the intersection of existing patches and source code into consideration to reduce the search space. CapGen [25] utilizes context information to prioritize patches. Empirically, its precision can reach 84% on four projects of Defects4J [25]. ssFix [21] leverages existing code that is syntax-related (i.e., structurally similar and conceptually related) to the context of a bug to produce patches for its repair. JAID [22] is designed based on detailed, state-based dynamic program analyses since grounding the repair generation and validation processes on rich state abstractions mitigates the overfitting problem. Elixir [27] can effectively synthesize patches from a repair space rich in method invocation expressions, by using a machine-learned model to rank the space of concrete repairs. AVATAR [29] is a pattern-based patch generation technique. It exploits fix patterns of static analysis violations as ingredients for generating candidate patches.

### B. Validation of APR-Generated Patches

Traditionally, the test cases associated with the buggy program under repair are used as the criteria for judging the correctness of APR techniques generated patches. GenProg, RSRepair, and AE reported to produce many correct patches under the assumption that a patch that passes the original test suite is regarded as correct. However, it has been shown in recent studies [32, 36] that this assumption does not hold true in practice due to the potential overfitting between the generated patches and the test suites.

Motivated by the above concern, recent studies employ new methods to assess patch correctness. One is to utilize **inde-**

**pendent test suites** generated by automatic generation tools. For example, Smith et al. [31] use general-purpose automatic test generation tool such as KLEE [43] to generate test suites for C language. For Java language, Xin et al. [34] propose DiffTGen, a test generation tool specially designed to generate tests that can identify incorrect patches generated by APR techniques. DiffTGen attempts to generate test cases that cover the syntactic and semantic differences between the generated patch and the developer-provided patch. A patch is labeled as incorrect if there exists a test case that exposes the differences in outputs of the programs. Another way is named **author annotation**, in which authors of the APR techniques manually check the generated patches and assess their correctness. This method is widely adopted by recent studies [20-23, 25-28].

However, a recent study [30] revealed that utilizing independent test suite can only detect a small part of incorrect patches when being used alone and author annotation suffers from subjectivity although it achieves high precision. Therefore, deeper analysis is desired to investigate the process of assessing patch correctness, especially, to understand how correct patches generated by APR techniques can achieve semantic equivalence while syntactic difference compared with those provided by developers. This study aims at bridging this gap.

III. STUDY DESIGN

In this section, we describe the details of our study design from three aspects: data selection, data filtering, and our research questions, respectively.

*A. Data Selection*

Our objective is to characterize and understand the correct patches (i.e., semantic equivalent compared with patches made by developers) that have been generated by existing studies, through which to provide guidance for future research towards author annotation of correct patches. To this end, we first present a Literature Review (LR) on the reported correct patches generated by recent techniques. We select Defects4J as our database since it is a widely used benchmark [20-29, 44, 45]. This benchmark contains 395 bugs extracted from six open source projects (i.e., JFreechart, Closure compiler, Apache commons-lang, Apache commons-math, Mockito, and Joda-Time). In particular, we study patches generated by 10 popular APR techniques (ACS [20], ssFix [21], JAID [22], CapGen [25], Elixir [27], SimFix [28], AVATAR [29], Nopol [42, 45], jGenProg [45], and jKali [45]). All these patches are labeled as correct after author annotation. Note that there are some other APR techniques for Java language, e.g., JFix [9] and NPEFix [46]. We exclude them from this study since they have not been evaluated on Defects4J. There are also some techniques that have been evaluated on Defects4J, e.g., SOFix [23] and SketchFix [26]. They are excluded since the generated patches are not publicly available. A recently proposed technique, ProbabilisticModel [24], uses a held-out test suite to assess patch correctness. However, as pointed out by the study [30] that using independent test suite alone is inefficient, the patches labeled as correct can potentially be incorrect. We thus exclude those patches generated by this technique from our study to avoid potential bias to our results. We also exclude the tool Cardumen [47] since the correctness of the generated patches is not labeled in the original evaluation of the tool. AVATAR provides three sets of patches generated under different assumptions of fault localization. The first one assumes that the perfect location, i.e., the faulty code elements, is known. The second one assumes the faulty method name is known and the third one makes no assumption on fault location. In this study, we take the first set of patches into consideration since AVATAR produces 34 correct patches under this assumption, more than those produced under the other two assumptions. We adopt such a heuristic with the aim to include more correct patches in our empirical study. The authors of AVATAR [29] introduced the concept named *Partially Fixed* which means a patch passes part of the failing test cases but not all of them. After a manual check, we find that six of the patches reported by AVATAR are, in fact, composed of several *partially fixed* patches. These *partially fixed* patches fix a multi-location bug collectively. Traditionally, a valid patch must contain all the necessary modifications required to repair the bug [2, 3, 39]. We thus do not take these *partially fixed* patches into consideration since they do not satisfy this requirement. After this step, we collect totally 185 patches.

*B. Sanity Check of Data*

The correctness of the collected 185 patches are annotated by the original authors of the APR tools. However, author annotation, as pointed by a recent study [30], is subjective and might produce false positives (incorrect patches but annotated as correct by the authors accidently). Specifically, Nopol generates a patch annotated as correct by the authors for Math#73 [45], however, this patch is proved to be incorrect in the study [30]. Therefore, it motivates us to have a sanity check manually over these collected patches to guarantee their correctness. A recent study [30] checked the correctness of 11 patches in our dataset and detects two false positives. In our study, we choose to adopt their conclusions and filter these two patches. For the left 174 patches, we further perform a sanity check to label their correctness aiming at filtering the potential false positives.

To perform our sanity check, we interview 27 undergraduate students in our college to judge whether patches generated by APR tools are semantically equivalent to the ground truth human patches. These students are in their third year and have completed several programming courses on Java. Although they are not experts, they possess the cognition about object-oriented programming and can easily understand Java program. We divide these students into nine groups and each patch is labeled by three participants, making each participant judge 19 or 20 patches in total. The design follows an existing study [30]. Specifically, for each patch, we provide the participants with the corresponding buggy program, the ground truth patch, an APR tool generated patch, the corresponding test files, and the detailed information about the failing test case. Based on this information, participants are asked to evaluate the correctness of the patch and they can choose one of the following options: "Yes", "No", or "I don't know". Each participant is required to finish the whole task in 2 hours and they can provide some comments about their decisions if they like. The results are shown in Table I. The number of patches in which all participants agree on each patch's label is 153 (87.9% of all patches), of which 151 patches are labeled as correct and 2 are labeled as incorrect. For 20 out of 174 patches (11.5% of all patches), there is a majority decision (i.e., not all participants but most participants agree on one label), out of which 16 and 4 patches are identified as correct and incorrect, respectively. We thus filter the 6 patches that are labeled as incorrect by two or three participants. For the remaining one patch which does not possess a majority decision, the authors decide to label it as correct after a discussion. Finally, 177 valid patches are collected as our study subject. The detailed information of our dataset is illustrated in Table II.

TABLE I. RESULTS OF PARTICIPANT ANNOTATIONS

|  | All Agree | Majority Agree | Total |
|---|---|---|---|
| Correct | 151 | 16 | 167 |
| Incorrect | 2 | 4 | 6 |
| Total | 153 | 20 | 173 |

TABLE II. DATASET OF COLLECTED PATCHES

| Tool | Chart | Closure | Lang | Math | Mockito | Time | Total |
|---|---|---|---|---|---|---|---|
| CapGen | 5 | 0 | 7 | 14 | 0 | 0 | 26 |
| SimFix | 4 | 6 | 8 | 14 | 0 | 1 | 33 |
| Nopol | 1 | 0 | 2 | 1 | 0 | 0 | 4 |
| jGenProg | 0 | 0 | 0 | 4 | 0 | 0 | 4 |
| jKali | 0 | 0 | 0 | 1 | 0 | 0 | 1 |
| JAID | 4 | 9 | 5 | 7 | 0 | 0 | 25 |
| Elixir | 4 | 0 | 8 | 12 | 0 | 2 | 26 |
| AVATAR | 5 | 7 | 4 | 6 | 2 | 2 | 26 |
| ssFix | 2 | 1 | 5 | 7 | 0 | 0 | 15 |
| ACS | 2 | 0 | 3 | 11 | 0 | 1 | 17 |
| Total | 27 | 23 | 42 | 77 | 2 | 6 | 177 |

*C. Research Questions*

In this study, we aim at investigating the following three research questions.

**RQ1** *How do machine-generated correct patches differ from developer-provided ones?* Machine-generated patches differ from each other mainly in two aspects: the edit point and the code modification. The edit point refers to the modified location while the code modification usually refers to the atomic operations conducted based on certain fixing ingredients. Fixing ingredients are those existing code elements reused in code modifications to generate patches [25]. Usually, there are three kinds of atomic operations: insertion, deletion, and replacement [2]. In this research question, we aim to study the differences between machine-generated correct patches and developer-provided patches from these two perspectives. In particular, two modifications are regarded as the same if both the atomic operation and the concerned fixing ingredients are the same. As a result, we manually divide our collected patches into the following four types according to these two aspects: Same Location Same Modification (SLSM) means the two patches are the same; Same Location Different Modification (SLDM) means machine-generated patch operates different modification at the same edit point compared with developer-provided one; Different Location Same Modification (DLSM) means the identical modification is performed at a different place in machine-generated patch; and Different Location Different Modification (DLDM) means different code modifications are performed in different locations.

**RQ2** *How do different types of patches distribute?* Our aim is to provide guidance for future author annotation. It is of great value for this study if we observe that a certain type of patches is generated by a certain type of APR techniques or in a certain project in Defects4J. Thus, we further study the distribution of different types of patches. Specifically, we measure distributions in terms of two perspectives (i.e., Defect4J projects and APR techniques).

**RQ3** *Do APR tools tend to generate correct patches but different from the developer-provided ones for bugs with certain characteristics?* We study two aspects of characteristics for bugs. One aspect is fixing complexity. We use metrics that have been analyzed in previous studies [44, 48] such as patch size and number of modified files for assessment. Our intuition is that if a bug is complex (i.e., the developer-provided patch is complex), APR techniques are likely to generate patches that are different from its ground truth since there might be multiple ways to correctly fix such a bug if it concerns multiple code elements. The other aspect is test adequacy of the buggy class. We use line coverage and branch coverage to measure it as existing studies do [49, 50]. Our intuition is that the test quality measured by line and branch coverages is related with the type of correct patches generated for this bug since existing studies have shown that the correctness (i.e., plausible, overfitting, or correct) of APR generated patches has strong correlation with the test quality [32, 34, 36]. If a certain characteristic can lead APR techniques to generate correct patches different from the ground truth, we can provide guidance for assessing the correctness of the patches generated for bugs with this feature.

## IV. RESULTS AND ANALYSES

In this section, we present the answers to our three research questions.

*A. RQ1: Patch Differences*

According to whether the edit location or the code modification is the same as the developer-provided patch, a machine-generated patch can be classified into four types: SLSM, SLDM, DLSM, and DLDM. We introduce the case for each type in the following.

*1) SLSM*

This type indicates that the machine-generated correct patch is identical to the developer-provided one. This type of patches is common, e.g., many APR techniques like SimFix and CapGen generate SLSM patches for bug Chart#1. Totally, there are 132 SLSM patches in our dataset.

*2) SLDM*

Fig. 1 shows the patch for Closure#115 generated by SimFix. In the developer-provided patch for this bug, lines 731-733 are deleted directly. In the machine-generated patch, it changes the condition in the if-statement in line 731. Note that if the program can execute to line 731, the parameter *cArg* must be non-empty. Thus, the condition in line 731 cannot be satisfied, which means the operation in line 732 will never be executed. This is semantically equivalent to directly deleting these statements and thus is a case of SLDM.

Totally, there are 33 SLDM patches. To investigate the differences between machine-generated correct patches and developer-provided ones, we review the Diff views to characterize their different operations over code elements. The develop-

```
730    //start of patch
731    if (cArg==null&&NodeUtil.canBeSideEffected(cArg)) {
732        return CanInlineResult.NO;
733    }
734    //end of patch
```

Fig. 1. The patch for Closure#115 generated by SimFix.

er-provided patch is used as the oracle and we check what changes the machine-generated patch makes such that it achieves semantic equivalence. This comparison is different from previous study [48] where the authors make correspondence between original buggy program (rather than APR generated patch) and its developer-provided patch. The results are shown in Table III. The naming rule is referred to the previous study [48] where the authors named the repair actions of patches in Defects4J. The column Category represents different types of code elements such as Assignment and Variable. The column Operation displays detailed operations which are composed by code elements and the corresponding actions like addition and removal. For example, the patch for Closure#115 generated by SimFix which is shown in Fig. 1 uses a conditional branch to finish the deletion operation in developer-provided patch. The generated patch contains an extra conditional branch, thus, the different operations between these two patches are classified into *conditional branch addition*. The third column shows the number of machined-generated patches under this kind of operation. In the last column, we list a sample of machine-generated patch for each operation by illustrating the APR technique followed by the bug ID.

The majority of SLDM patches involve code elements related to conditional block (21/33). Machine-generated correct patches may add or remove a conditional branch or change (i.e., modify, expand, or reduce) the content in the conditional expression compared to developer-provided patches. Machine-generated correct patches may also perform operations on method calls such as adding method calls or calling different methods. Modifying assignments or variables may help to achieve semantic equivalence but their frequencies of occurrence are rather low. Note that the category Statement does not appear in the study [48]. In our observation, this category contains three samples of incomplete statement removal which means the machine-generated patch deletes some but not all the statements deleted by developer-provided patch.

*3) DLSM*

Fig. 2 shows the patch for Math#53 generated by CapGen. In the machine-generated patch, it inserts an if-statement and the corresponding operations as the developer-provided one does. The difference is that the insertion point in this patch is at the beginning of this function while in developer-provided one, it is under line 8. Note that the method call in line 8 only checks if the object *rhs* is null and throws an exception if it is. Thus, the order of these two statements does not influence the program. This is a case of DLSM.

During our study, we observe 2 DLSM patches in total. To further observe where the operations in machine-generated patches are conducted, we investigate from two scopes. *Same Method* means the two edit points in the machine-generated and developer-provided patches are in the same method while *Same Class* means the two edit points are not in the same method but in the same class. Results in Table IV show that all the edit points of these two DLSM patches are in the same methods as those of the corresponding developer-provided pat-

TABLE III. DETAILS OF DIFFERENCES BETWEEN SLDM PATCHES AND DEVELOPER-PROVIDED PATCHES

| Category | Operation | #patches | Sample |
|---|---|---|---|
| Conditional | Conditional Expression Modification | 5 | JAID-Lang#33 |
| | Conditional Expression Expansion | 3 | SimFix-Math#63 |
| | Conditional Expression Reduction | 1 | Nopol-Lang#58 |
| | Conditional Branch Removal | 4 | ssFix-Lang#33 |
| | Conditional Branch Addition | 8 | SimFix-Closure#115 |
| Method Call | Method Call Addition | 3 | Elixir-Time#15 |
| | Method Call Replacement | 4 | CapGen-Lang#57 |
| Assignment | Assignment Modification | 1 | ssFix-Math#80 |
| Variable | Variable Modification | 1 | JAID-Chart#24 |
| Statement | Incomplete Statement Removal | 3 | JAID-Closure#40 |

```
 5     public Complex add(Complex rhs)
 6         throws NullArgumentException {
 7   +     if ((isNaN) || (rhs.isNAN)) { return org.apache.commons.math.complex.Complex.NaN;}
 8         MathUtils.checkNotNull(rhs);
 9         return createComplex(real + rhs.getReal(),
10             imaginary + rhs.getImaginary());
11     }
```

Fig. 2. The patch for Math#53 generated by CapGen.

TABLE IV. DETAILS OF DIFFERENCES BETWEEN DLSM PATCHES AND DEVELOPER-PROVIDED PATCHES

| | Same Method | Same Class |
|---|---|---|
| #Patches | 2 | 0 |

ches.

*4) DLDM*

In Fig. 3, we illustrate the developer-provided and machine-generated correct patches for Chart#3 generated by SimFix.

The developer-provided one assigns two attributes of the object named *copy* under the function named *createCopy*. In the following statements of this function, *copy* will call another function named *add*, where the modification of SimFix-generated patch occurs. The machine-generated patch calls another function named *findBoundsByIteration* in the function *add*. In the function *findBoundsByIteration*, there are two statements assigning the same values to the attributes of the object as the developer-provided patch does. Note that there is a conditional statement under the end of machine-generated patch. If the program executes to this point without throwing an exception, the parameter *added* is always true, which means the method *updateBoundsForAddedItem* will be called and it is exactly what the loop in the method *findBoundsByIteration* does. The method *updateBoundsForAddedItem* does not change the object *item*, thus, the machine-generated correct patch is semantic equivalent to the developer-provided one, being a case of DLDM.

To further illustrate the differences between DLDM patches and their corresponding developer-provided ones, we investigate from the aforementioned two aspects: the code operation and the edit point. We summarize the situations in Table V. Totally, ten patches are recognized as DLDM in our study, seven of which conduct operations in the same methods as their corresponding developer-provided patches. The different code operations are all about conditional block and method call, which is consistent with the results in Table III where these two types account for the majority. From our results, patches contain different edit points from the developer-provided ones (DLSM and DLDM patches) perform code modifications at le-

```
1056     TimeSeries copy = (TimeSeries) super.clone();
1057  +  copy.minY = Double.NaN;
1058  +  copy.maxY = Double.NaN;
1059     copy.data = new java.util.ArrayList();
```

a) Developer-provided patch

```
//start of patch
findBoundsByIteration();
//end of patch
if (added) {
    updateBoundsForAddedItem(item);
    //check if this addition will exceed the maximum item count...
    if (getItemCount() > this.maximumItemCount) {
        TimeSeriesDataItem d = (TimeSeriesDataItem) this.data.remove(0);
        updateBoundsForRemovedItem(d);
    }
}
```

b) Machine-generated patch

```
private void findBoundsByIteration() {
    this.minY = Double.NaN;
    this.maxY = Double.NaN;
    Iterator iterator = this.data.iterator();
    while (iterator.hasNext()) {
        TimeSeriesDataItem item = (TimeSeriesDataItem) iterator.next();
        updateBoundsForAddedItem(item);
    }
}
```

c) The function named *findBoundsByIteration*

Fig. 3. Patches for Chart#3.

TABLE V. DETAILS OF DIFFERENCES BETWEEN DLDM PATCHES AND DEVELOPER-PROVIDED PATCHES

| Category | Operation | Same Method | Same Class |
|---|---|---|---|
| Conditional | Conditional Expression Modification | 3 | 0 |
| | Conditional Expression Reduction | 1 | 0 |
| | Conditional Branch Addition | 2 | 0 |
| Method Call | Method Call Addition | 1 | 1 |
| | Method Call Replacement | 0 | 2 |

TABLE VI. STATISTICS OF EDIT POINT DISTANCES OF DLDM PATCHES

| | Min | Median | Mean | Max |
|---|---|---|---|---|
| Edit Point Distance | 1 | 24 | 87.5 | 430 |

ast in the same class. To measure the distance between the edit points of machine-generated and developer-provided patches, we define the concept of edit point distance, which is the number of lines between the two points in the program. For the two DLSM patches, the edit point distances are both 1 while the statistics of the edit point distances of DLDM patches are illustrated in Table VI. The minimum in Table VI occurs in the patch for Chart#26 generated by JAID. It adds another conditional statement at one line before the edit point in the ground truth. The median and mean of edit point distances of DLDM patches can reach tens of lines of code. The maximum occurs in the patch for Chart#3 generated by SimFix as shown in Fig. 3. SimFix generates a patch which performs method call addition operation under another method in the same class, making the edit point distance up to 430. To recap, the edit point distances of DLDM patches are much larger then those of DLSM patches.

Totally, we observe 132 SLSM patches, 33 SLDM patches, 2 DLSM patches, and 10 DLDM patches, from which two findings can be concluded. First, 74.6% (132/177) of the patches belong to SLSM type, which means that a large part of the correct patches are identical to the developer-provided ones.

This indicates the great progress of the repair ability of APR techniques developed in recent years. A previous study [32] shows that a large part of patches generated by GenProg, RSRepair, and AE are semantically equivalent to functionality deletions, while we find that a large part of the patches generated by recent APR techniques are the same as developer-provided ones. Second, although a large amount of the patches are the same as developer-provided ones, there are still over 25% patches being different from the ground truth. This shows that APR techniques do not need to generate patches exactly like how the developers do. Among these patches, more than 95% (43/45) are SLDM or DLDM patches, which means they perform different code modifications compared with developer-provided patches. Our finding indicates that we do not need to be afraid of generating code different from the ground truth, reflecting the view of Monperrus in [41].

> **RQ1.** APR-generated correct patches can be classified into four types based on their edit points and code modifications, while most of them (around 75%) are identical to their ground truth (i.e., SLSM patches).

### B. Patch Distribution

In this research question, we investigate the distribution of different types of patches from two perspectives (Defects4J projects and APR techniques). Specifically, we investigate how many different types of patches are generated by existing APR techniques for bugs in each project and how many different types of patches are generated by each APR technique. Our aim is to observe the characteristics of correct patches from a more detailed perspective, e.g., if a certain type of correct patches is only generated by a particular APR technique or for bugs in a particular project. Table VII and Table VIII show such distributions respectively.

In Table VII, the first column indicates the six projects in the Defects4J benchmark. The following four columns indicate the number of different types of correct patches that have already been generated for bugs from each project. The last column indicates the total amount of correct patches generated for bugs from each project. In Table VIII, the first column indicates the ten APR techniques considered by our study. The following four columns indicate the number of different types of correct patches generated by each technique. The last column indicates the total number of correct patches generated by each technique. Please note that the numbers shown in the last column in Table VIII may differ from the numbers of correctly-fixed bugs shown in the corresponding original papers. That is because a tool may generate multiple correct patches for a certain bug and we take all these patches into consideration. For example, CapGen correctly fixes 22 bugs but generates multiple correct patches for 4 bugs. Thus, the corresponding number shown in Table VIII is 28. From the results, we can observe the following findings.

**From the perspective of Defects4J projects.** From Table VII, 85.7% (66/77) of the patches generated for project Math are SLSM patches, which is the highest among the six projects. DLSM patches are generated only in two projects (i.e., Chart and Math) and DLDM patches are generated only in three projects (Chart, Lang, and Math). Besides, fault localization is crit-ical for fixing bugs in Closure, Mockito, and Time, since all the correct patches are generated based on the correct buggy location (i.e., only SLSM and SLDM patches have been generated).

TABLE VII. PATCH DISTRIBUTION FROM DEFECTS4J PROJECTS PERSPECTIVE

| Project | #SLSM | #SLDM | #DLSM | #DLDM | Total |
|---|---|---|---|---|---|
| Chart | 22 | 1 | 1 | 3 | 27 |
| Closure | 17 | 6 | 0 | 0 | 23 |
| Lang | 24 | 12 | 0 | 6 | 42 |
| Math | 66 | 9 | 1 | 1 | 77 |
| Mockito | 0 | 2 | 0 | 0 | 2 |
| Time | 3 | 3 | 0 | 0 | 6 |
| Total | 132 | 33 | 2 | 10 | 177 |

TABLE VIII. PATCH DISTRIBUTION FROM APR TECHNIQUES PERSPECTIVE

| Technique | #SLSM | #SLDM | #DLSM | #DLDM | Total |
|---|---|---|---|---|---|
| CapGen | 22 | 2 | 2 | 0 | 26 |
| SimFix | 23 | 6 | 0 | 4 | 33 |
| AVATAR | 18 | 8 | 0 | 0 | 26 |
| Nopol | 0 | 1 | 0 | 3 | 4 |
| jGenProg | 4 | 0 | 0 | 0 | 4 |
| jKali | 1 | 0 | 0 | 0 | 1 |
| JAID | 14 | 9 | 0 | 2 | 25 |
| Elixir | 22 | 4 | 0 | 0 | 26 |
| ACS | 16 | 0 | 0 | 1 | 17 |
| ssFix | 12 | 3 | 0 | 0 | 15 |
| Total | 132 | 33 | 2 | 10 | 177 |

**From the perspective of APR technique.** From Table VIII, DLSM patches are generated only by CapGen. This is probably because CapGen searches over all potential buggy points to generate patches. The tools jGenProg and jKali can only generate SLSM patches while CapGen, SimFix, and JAID can generate three kinds of patches. The DLDM patches can be generated by only four techniques (i.e., SimFix, Nopol, JAID, and ACS).

**The projects with a small number of patches.** While each of the other four projects contains more than 20 patches, the number of patches for Mockito and Time is only 8. This is, however, caused by different reasons. For Mockito, it is due to the neglect of developers of APR techniques during the evaluation [44]. This phenomenon calls for a more comprehensive evaluation for APR techniques as the authors in [44] argued. On the contrary, all the recent APR techniques have been evaluated on the Time project due to the statistics in the study [44]. It is potentially caused by the low repair ability of the state-of-the-art APR techniques. Thus, it calls for more actions towards repairing bugs in Time.

> **RQ2.** The distribution of APR-generated correct patches diverges for the aspects of Defects4J projects and APR techniques: most of the patches (around 85%) generated for project Math are SLSM patches while DLSM patches are only generated by CapGen.

## C. Bug Characteristics

In this research question, we aim to investigate whether APR tools tend to generate correct patches but different from the developer-provided ones for bugs with certain characteristics. Specifically, we investigate this question from two aspects, which are patch complexity and test adequacy inspired by the following existing studies [44, 51-57].

Substantial studies have been proposed to characterize patch complexity. A study of Linux Kernel patching process [51] measures *locality of patches* through three indicators (i.e., files, hunks, and lines). Another previous study [52] annotates the Defects4J bugs with patch size and number of modified files to compute the complexity. A more recent study [44] has performed detailed analysis of patch characteristics in Mockito project from Defects4J. We choose the four indicators as they did. It is widely known that there are three types of code changes: addition, deletion, and modification. Addition and deletion appear as lines of codes are added or deleted consecutively or separately in source code. Modification appears as sequences of removed lines are straightly followed by added lines or vice-versa. The patch size is the sum of the number of lines of these three types of code changes in the patch. Composed by the combination of addition, deletion, and modification of lines, a chunk is a sequence of continuous changes in a file. The number of chunks of a patch can provide insights on how a patch is spread through the source code and further give information about how complex the patch is: the more chunks means the more buggy points in the program, and thus the more complexly for fixing this bug. Similarly, the number of modified files and the number of modified methods are also two important indicators. The larger they are, the more program elements are involved in the patch, and the more complex the patch is. These four indicators have been shown to represent patch complexity well [44, 52, 53].

Statement coverage and branch coverage are widely-used in debugging tasks to represent the test adequacy [54-57]. Specifically, statement coverage is used to calculate and measure the number of statements in the source code which have been executed while branch coverage is used to calculate and measure the number of reachable branches in the Control Flow Graph of the program which have been executed.

We conduct the statistics on the six characteristics mentioned above of the bugs for which at least a correct patch is generated. The data of patch complexity is from the previous study [48] in which the characteristics of each bug in Defects4J have been analyzed. The data of test adequacy is calculated by Cobertura[1] which is a free Java tool being widely-used in recent studies [49, 50]. If different types of patches are generated for the same bug, the data of this bug is added into all the relevant types for analysis. The distributions of the six characteristics on bugs for which different types of patches are generated are illustrated in Fig. 4. We also conduct a significant difference test to check if the differences in the distributions are statistically significant. For each characteristic, we consider the data of bugs for which SLSM patches are generated as standard and make comparison between them and the data of bugs for which SLDM and DLDM patches are generated using Mann-Whitney-Wilcoxon test. The intuition is that we aim to investi-gate whether bugs for which patches different from the ground truth are generated possess obvious characteristics, compared with bugs for which SLSM patches are generated. We ignore the p-values of bugs for which DLSM patches are generated since there are only two bugs. Therefore, the data is not enough for a significant difference test. The results are shown in Table IX with the average value of each characteristic for each type of bug.

---
[1] http://cobertura.github.io/cobertura

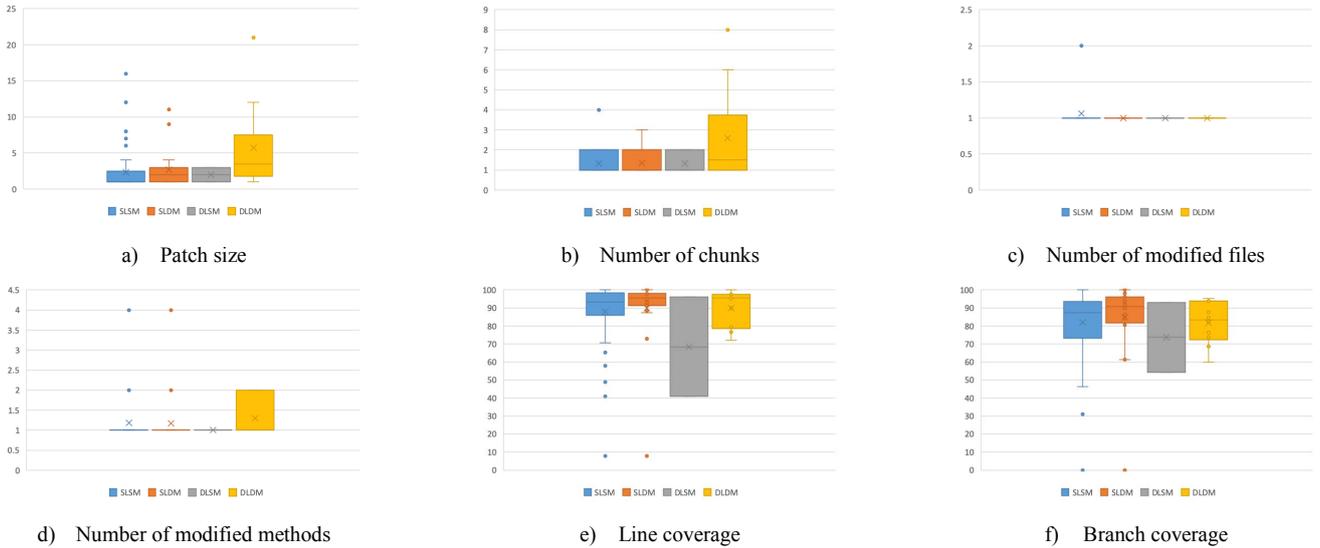

Fig. 4. Distributions of Bug Characteristics

TABLE IX. RESULTS OF THE SIGNIFICANT DIFFERENCE TEST

| Characteristic | SLSM | SLDM | | DLSM | | DLDM | |
|---|---|---|---|---|---|---|---|
| | ave | ave | p-v | ave | p-v | ave | p-v |
| patch size | 2.32 | 2.70 | 0.183 | 2 | - | 5.7 | 0.008 |
| number of chunks | 1.32 | 1.35 | 0.862 | 1 | - | 2.6 | 0.073 |
| number of modified files | 1.06 | 1 | 0.226 | 1 | - | 1 | 0.423 |
| number of modified methods | 1.18 | 1.17 | 0.463 | 1 | - | 1.3 | 0.275 |
| line coverage | 87.88% | 89.84% | 0.406 | 68.5% | - | 90.04% | 1 |
| branch coverage | 81.86% | 84.69% | 0.266 | 73.65% | - | 81.62% | 0.559 |

The previous study [48] shows that most of the patches modify only one file in Defects4J, thus the difference over the number of modified files indicator is not significant, as is shown in Table IX. The average values of the four types of bugs are close.

For bugs for which SLDM patches are generated, their averages in patch size, number of chunks, and line and branch coverages exceed those of the bugs for which SLSM patches are generated and their average in number of modified methods is only a little less than that of the SLSM. These differences are all insignificant since all these p-values are higher than 0.05.

For bugs for which DLSM patches are generated, their averages are lower than those of the SLSM in all the six characteristics, especially in line and branch coverages where the averages of the other three types of bugs are higher than 80% while the averages of this type are only about 70%.

For bugs for which DLDM patches are generated, their averages in patch size and number of chunks are much higher than those of other three types of bugs (5.7 and 2.6, respectively). The test results indicate that the difference between this type of bugs and bugs for which SLSM patches are generated is significant in patch size with the p-value reaching 0.008. This type of bugs also possesses the highest line coverage average, but the difference is insignificant. The other three averages are close to those of the SLSM.

For bug Lang#57, the line and branch coverages of the buggy class are 7.9% and 0, respectively, making it an outlier. This happens because the raised exception in the *setup* method causes the failing tests abort before the execution of the code in the class under test.

Our results indicate that on one hand, APR techniques are prone to generate DLDM patches for bugs which are complex for repairing since the difference in the patch size is significant. On the other hand, APR techniques have generated DLSM patches for bugs whose test adequacies are lower than other bugs.

> **RQ3.** APR techniques are prone to generate DLDM patches for bugs with large patch sizes. They also generate DLSM patches for bugs with low test adequacies but the difference significance cannot be measured.

## V. DISCUSSION

In this section, we first provide implications of our findings. We then discuss potential threats to validity.

### A. Implications

To recap, we have obtained empirical results via investigating patch differences, distributions, and characteristics. Based on these results, we distill several implications as follows.

- On the evaluation of patches generated by synthesize-based APR techniques

General program repair techniques can typically be divided into two main branches: search- and synthesize-based repair methods. Search-based repair methods generate patch candidates by searching within a predefined fault space determined by Fault Localization (FL) techniques and then validate these candidates against the provided test suite. Synthesize-based repair methods, on the contrary, utilize semantic information to synthesize patches. Among the ten APR techniques investigated in this study, only Nopol and ACS are synthesize-based, but they totally generate nearly half (4/10) of the DLDM patches according to the results shown in Table VIII. This implicates that synthesize-based tools are prone to generate correct patches that are syntactically different from developer-provided ones, especially for Nopol where no SLSM patches are generated. This implicates that comprehensive analysis should be conducted when evaluating the correctness of patches generated by future synthesize-based tools, and patches should not be casually labeled as incorrect simply because they are not syntactically the same as the developer-provided ones.

- On the location of edit points

Although correct patches can be generated at different places compared with their ground truth and the edit point distances could even exceed one hundred lines (cf. Tables IV and V), corrected patches are more likely to be generated when the edit distances are smaller. For example, 95.5% (169/177) of the correct patches are generated within distance of 1, and that ratio is only 3.4% (6/177) when the distance exceeding 10. This reflects the importance of fault localization in generating correct patches. Another interesting finding revealed by Table V is that all the edit points of correct patches concerning conditional block are generated within the buggy methods. This indicates that patches that deal with the conditional blocks should focus on the conditional blocks in the buggy method.

- On the importance of conditional block

According to our results, the code modifications of 63.6% (21/33) of SLDM patches and 60% (6/10) of DLDM patches are related to conditional block. This reflects the necessity of APR techniques such as ACS and Nopol that are designed specific for conditional blocks and calls for more in-depth research towards this direction. Besides, among the five different types of operations in the Conditional category, *conditional block addition* and *conditional expression modification* are the most popular ones (cf. Tables III and V). This indicates that machine-generated patches that make certain adjustments on conditional blocks compared with the ground truth may be correct. However, the correctness cannot be fully guaranteed since four out of seven patches that have been filtered out in our sanity check also belong to this code modification category (two are *conditional expression modification* and two are *conditional expression expansion*).

- On the assessment of DLSM patches

The only two correct DLSM patches are generated by CapGen (cf. Table VIII). These two patches perform the same code modification with the ground truth but at different points, achieving semantic equivalence. However, another two DLSM patches generated by CapGen were filtered during our sanity check. These patches perform code modification outside of a conditional branch, but the same code modification is performed within the conditional branch by developers. Our participants consider these two patches as false positives since they affect the control flow of the program. Thus, we should check carefully about the control and data flow of the program when assessing the correctness of DLSM patches in the future.

- On the substantial portion of method calls

Machine-generated patches also prefer to utilize *method call addition* and *method call replacement* to fulfill semantic equivalence (cf. Tables III and V). The edit points of these patches are not even restricted to the buggy method: three of their edit points reside outside of the buggy method. This means that in programs with complex logical structures, different method invocations may achieve the same target. For instance, in the patch generated by ACS for bug Chart#7, the method *trim()* is not called while this method is called in the developer's patch. Thus, we should analyze the program comprehensively when facing patches that contain code modifications about method calls.

- On the bugs for which patches different from the ground truth have been generated

We list 35 bugs for which correct patches that are different from the developer-provided ones have been generated in Table X. We call for attention for evaluating patches for these bugs since our empirical study reveals that there is not only one way to fix them.

TABLE X. BUGS THAT NEED ATTENTION

| Project | Bug ID |
| --- | --- |
| Chart | 3; 5; 11; 24; 26 |
| Closure | 2; 33; 38; 40; 115; 126 |
| Lang | 7; 10; 16; 26; 33; 39; 41; 43; 44; 50; 51; 55; 57; 58 |
| Math | 32; 35; 50; 53; 63; 80 |
| Mockito | 29; 38 |
| Time | 7; 15 |

*B. Threats to Validity*

The main threats to the validity of our results belong to the internal and external validity threat categories.

Internal validity threat corresponds to the dataset in our study. Author annotation unavoidably suffers from bias. To fill this gap, we conduct a sanity check by asking undergraduate students to judge the correctness of those patches annotated by the original authors. These students have no actual development experience in industry. As a result, whether they can filter all the false positives is questionable. However, this threat is limited since 1) we believe that considerable effort has been made by authors to ensure the quality of their labels; 2) the previous study [30] shows that the possibility for authors to generate wrong labels is rather low; 3) we do not find particularly obvious mistakes during our investigation; and 4) each collected patch is labeled by three individual participants. Since the majority (74.6%) of our collected patches are SLSM patches which are the same to the ground truth, the time limitation (2 hours) we set for this interview is sufficient for the participants to judge non-SLSM patches in their own task.

External validity threats correspond to the generalization of our results. Due to the unavailability of patches generated by APR techniques such as SOFix and SketchFix, we exclude them from our dataset. Thus, it is possible that results may differ in these patches. We select Defects4J dataset as our benchmark, as a result, bugs from other databases such as BEARS [58] and Bugs.jar [59] are neglected. Patches generated for these bugs may demonstrate different characteristics. The threat is limited when considering the popularity of Defects4J for being the evaluation criterion of recent studies [20-29].

## VI. RELATED WORK

In Section II, we have described some popular APR techniques and methods for patch evaluation. In this section, we introduce some empirical studies on patch correctness assessment and biases in software engineering.

### A. Patch Correctness Assessment

Qi et al. [32] empirically studied patches generated by GenProg, RSRepair, and AE. They found that the presented evaluations of these techniques suffer from the fact that the testing infrastructure used to validate the candidate patches contains errors that cause the systems to incorrectly accept implausible patches that do not even pass all the test cases in the validation test suite. They subsequently corrected these errors and found that 1) the systems generate much more plausible patches than correct patches and 2) the majority of the plausible patches, including all correct patches, are equivalent to a single modification that deletes functionality. They then presented a novel automatic patch generation system, Kali, that works only with simple patches that delete functionality. The experimental results showed that Kali generates at least as many correct patches as prior techniques (GenProg, RSRepair, and AE). Smith et al. [31] evaluated two repair tools (GenProg and RSRepair) on a publicly available benchmark of 998 bugs. They used two test suites per program: one is training data used to construct a patch, and the other is evaluation data used to evaluate the quality of the patch. They found that 1) GenProg and RSRepair are less likely to repair programs that fail more training tests, 2) patches that are overfitting to the training test suite often break undertested functionality, and 3) higher coverage test suites lead to higher quality patches. Xin et al. [34] proposed DiffTGen which identifies a patched program to be overfitting by first generating new test inputs that uncover semantic differences between the original faulty program and the patched program, then testing the patched program based on the semantic differences, and finally generating test cases. They further showed that an automatic repair technique, if configured with DiffTGen, could avoid yielding overfitting patches and potentially produce correct ones. Xiong et al. [33] proposed a novel approach that heuristically determines the correctness of the generated patches to reduce the number of incorrect patches generated. Their core idea is to exploit the behavior similarity of test case executions. Empirically, their approach successfully prevented 56.3% of the incorrect patches to be generated when being evaluated on a dataset consisting 139 patches generated from 5 APR techniques, without blocking any correct patches. In a more recent study, Le et al. [30] assessed reliability of author and automated annotations on patch correctness assessment. They first constructed a gold set of correctness labels for 189 patches through a user study and then compared labels generated by author and automated annotations with this gold set to assess reliability. They found that although independent test suite alone should not be used to evaluate the effectiveness of APR, it can be used to augment author annotation. Yu et al. [60] studied the feasibility of using automatic test generation to alleviate patch overfitting. They divided the overfitting problem into two classifications (i.e., regression introduction and incomplete fixing) and found automatic test generation is effective in alleviating regression introduction. Our study is different from the mentioned studies in that we objectively summarize all the available correct patches and conduct a detailed analysis, aiming at providing insightful experience for future author annotation.

### B. Biases in Software Engineering

A number of empirical studies have analyzed biases issues that affect how software engineering solutions are evaluated. Liu et al. [17] identified and investigated a practical bias caused by the fault localization (FL) step in a repair pipeline. Their main findings included 1) only a subset of Defects4J bugs can be currently localized by commonly-used FL techniques and 2) current practice of comparing state-of-the-art APR systems is potentially misleading due to the bias of FL configurations. Tu et al. [61] investigated the data leakage, which results from ignoring the chronological order in which the data were produced. They examined existing literature and confirmed that 11 out of 58 studies have leakage problem. They further recommended researchers and practitioners who attempt to utilize issue tracking data to have a full understanding of the origin and change of the data. Wang et al. [44] investigated the bias caused by the evaluation process in APR. They recommended that more bugs should be considered to avoid the potential overfitting and make the conclusion more generalized. Rodriguez et al. [62] studied reproducibility in Empirical Software Engineering (ESE) by investigating how it has been addressed in studies where SZZ, a widely-used algorithm to detect the origin of a bug, has been applied. They confirmed that reproducibility is not commonly found and recommended to take reproducibility and other related aspects into consideration to increase the credibility of the research results. The goal of our study is similar to the mentioned studies since we want to reduce bias in the author annotation process in APR pipeline.

## VII. CONCLUSION AND FUTURE WORK

In this study, we conducted a dissection on the correct patches generated by the state-of-the-art APR techniques. We investigated the differences between these patches with their corresponding developer-provided patches and divided these patches into four types based on their edit points as well as code modifications. We then studied the distributions of these patches and the correlation between the characteristics of a bug and patches generated for the bug. We find that 1) APR techniques can generate patches that are different from the ground truth; 2) machine-generated correct patches can be divided into four types according to the edit points and code modifications: SLSM, SLDM, DLSM, and DLDM; and 3) APR techniques are more likely to generated DLDM patches for bugs which contain large patch sizes. Through our study, we confirm some opinions from previous studies (e.g., the view of Monperrus in [41]) and highlight several implications for future study about machine-generated patch correctness assessment.

In the future, we plan to expand our dataset to consider patches generated for other benchmarks such as Bugs.jar [58] to make our study more comprehensive.

## ACKNOWLEDGEMENT

This work was supported by the National Natural Science Foundation of China under Grant 61672529. All the data of this paper are publicly available at:
https://github.com/Kaka727/correct_patch_analysis

## REFERENCES

[1] Britton T, Jeng L, Carver G, et al. Reversible debugging software[J]. Judge Bus. School, Univ. Cambridge, Cambridge, UK, Tech. Rep, 2013.

[2] Weimer W, Nguyen T V, Le Goues C, et al. Automatically finding patches using genetic programming[C]//Proceedings of the 31st International Conference on Software Engineering. IEEE Computer Society, 2009: 364-374.

[3] Qi Y, Mao X, Lei Y, et al. The strength of random search on automated program repair[C]//Proceedings of the 36th International Conference on Software Engineering. ACM, 2014: 254-265.

[4] Long F, Rinard M. Automatic patch generation by learning correct code[C]//ACM SIGPLAN Notices. ACM, 2016, 51(1): 298-312.

[5] Ke Y, Stolee K T, Le Goues C, et al. Repairing programs with semantic code search (t)[C]//2015 30th IEEE/ACM International Conference on Automated Software Engineering (ASE). IEEE, 2015: 295-306.

[6] Nguyen H D T, Qi D, Roychoudhury A, et al. Semfix: Program repair via semantic analysis[C]//Software Engineering (ICSE), 2013 35th International Conference on. IEEE, 2013: 772-781.

[7] Mechtaev S, Yi J, Roychoudhury A. Angelix: Scalable multiline program patch synthesis via symbolic analysis[C]//Proceedings of the 38th international conference on software engineering. ACM, 2016: 691-701.

[8] Le X B D, Chu D H, Lo D, et al. S3: syntax-and semantic-guided repair synthesis via programming by examples[C]//Proceedings of the 2017 11th Joint Meeting on Foundations of Software Engineering. ACM, 2017: 593-604.

[9] Le X B D, Chu D H, Lo D, et al. JFIX: semantics-based repair of Java programs via symbolic PathFinder[C]//Proceedings of the 26th ACM SIGSOFT International Symposium on Software Testing and Analysis. ACM, 2017: 376-379.

[10] Xuan J, Monperrus M. Learning to combine multiple ranking metrics for fault localization[C]//2014 IEEE International Conference on Software Maintenance and Evolution. IEEE, 2014: 191-200.

[11] Wong W E, Gao R, Li Y, et al. A survey on software fault localization[J]. IEEE Transactions on Software Engineering, 2016, 42(8): 707-740.

[12] Zhang Z, Chan W K, Tse T H, et al. Non-parametric statistical fault localization[J]. Journal of Systems and Software, 2011, 84(6): 885-905.

[13] Abreu R, Zoeteweij P, Van Gemund A J C. On the accuracy of spectrum-based fault localization[C]//Testing: Academic and Industrial Conference Practice and Research Techniques-MUTATION (TAICPART-MUTATION 2007). IEEE, 2007: 89-98.

[14] Campos J, Riboira A, Perez A, et al. Gzoltar: an eclipse plug-in for testing and debugging[C]//Proceedings of the 27th IEEE/ACM International Conference on Automated Software Engineering. ACM, 2012: 378-381.

[15] Yang D, Qi Y, Mao X. An empirical study on the usage of fault localization in automated program repair[C]//Software Maintenance and Evolution (ICSME), 2017 IEEE International Conference on. IEEE, 2017: 504-508.

[16] Qi Y, Mao X, Lei Y, et al. Using automated program repair for evaluating the effectiveness of fault localization techniques[C]//Proceedings of the 2013 International Symposium on Software Testing and Analysis. ACM, 2013: 191-201.

[17] Liu K, Koyuncu A, Bissyande T, et al. You cannot fix what you cannot find! An Investigation of fault localization bias in benchmarking automated program repair[C]// In: Proceedings of IEEE International Conference on Software Testing, Validation, and Verification (ICST). IEEE, 2019.

[18] Zou D, Liang J, Xiong Y, et al. An Empirical Study of Fault Localization Families and Their Combinations[J]. IEEE Transactions on Software Engineering, 2019.

[19] Xiao Y, Keung J, Bennin K E, et al. Machine translation-based bug localization technique for bridging lexical gap[J]. Information and Software Technology, 2018, 99: 58-61.

[20] Xiong Y, Wang J, Yan R, et al. Precise condition synthesis for program repair[C]//Proceedings of the 39th International Conference on Software Engineering. IEEE Press, 2017: 416-426.

[21] Xin Q, Reiss S P. Leveraging syntax-related code for automated program repair[C]//Proceedings of the 32nd IEEE/ACM International Conference on Automated Software Engineering. IEEE Press, 2017: 660-670.

[22] Chen L, Pei Y, Furia C A. Contract-based program repair without the contracts[C]//Automated Software Engineering (ASE), 2017 32nd IEEE/ACM International Conference on. IEEE, 2017: 637-647.

[23] Liu X, Zhong H. Mining stackoverflow for program repair[C]//2018 IEEE 25th International Conference on Software Analysis, Evolution and Reengineering (SANER). IEEE, 2018: 118-129.

[24] Soto M, Le Goues C. Using a probabilistic model to predict bug fixes[C]//2018 IEEE 25th International Conference on Software Analysis, Evolution and Reengineering (SANER). IEEE, 2018: 221-231.

[25] Wen M, Chen J, Wu R, et al. Context-Aware Patch Generation for Better Automated Program Repair[C]// Proceedings of the 40th International Conference on Software Engineering. ACM, 2018.

[26] Hua J, Zhang M, Wang K, et al. Towards practical program repair with on-demand candidate generation[C]//Proceedings of the 40th International Conference on Software Engineering. ACM, 2018: 12-23.

[27] Saha R K, Yoshida H, Prasad M R, et al. Elixir: an automated repair tool for Java programs[C]//Proceedings of the 40th International Conference on Software Engineering: Companion Proceeedings. ACM, 2018: 77-80.

[28] Jiang J, Xiong Y, Zhang H, et al. Shaping Program Repair Space with Existing Patches and Similar Code[C]// The International Symposium on Software Testing and Analysis. 2018.

[29] Liu K, Koyuncu A, Kim D, et al. AVATAR: Fixing Semantic Bugs with Fix Patterns of Static Analysis Violations[C]//2019 IEEE 26th International Conference on Software Analysis, Evolution and Reengineering (SANER). IEEE, 2019: 456-467.

[30] Le X B D, Bao L, Lo D, et al. On Reliability of Patch Correctness Assessment[C]//Proceedings of the 41st International Conference on Software Engineering. IEEE, 2019.

[31] Smith E K, Barr E T, Le Goues C, et al. Is the cure worse than the disease? overfitting in automated program repair[C]//Proceedings of the 2015 10th Joint Meeting on Foundations of Software Engineering. ACM, 2015: 532-543.

[32] Qi Z, Long F, Achour S, et al. An analysis of patch plausibility and correctness for generate-and-validate patch generation systems[C]//Proceedings of the 2015 International Symposium on Software Testing and Analysis. ACM, 2015: 24-36.

[33] Xiong Y, Liu X, Zeng M, et al. Identifying patch correctness in test-based program repair[C]//Proceedings of the 40th International Conference on Software Engineering. ACM, 2018: 789-799.

[34] Xin Q, Reiss S P. Identifying test-suite-overfitted patches through test case generation[C]//Proceedings of the 26th ACM SIGSOFT International Symposium on Software Testing and Analysis. ACM, 2017: 226-236.

[35] Le Goues C, Dewey-Vogt M, Forrest S, et al. A systematic study of automated program repair: Fixing 55 out of 105 bugs for $8 each[C]//2012 34th International Conference on Software Engineering (ICSE). IEEE, 2012: 3-13.

[36] Long F, Rinard M. An analysis of the search spaces for generate and validate patch generation systems[C]//2016 IEEE/ACM 38th International Conference on Software Engineering (ICSE). IEEE, 2016: 702-713.

[37] Le X B D, Lo D, Le Goues C. Empirical study on synthesis engines for semantics-based program repair[C]//2016 IEEE International Conference on Software Maintenance and Evolution (ICSME). IEEE, 2016: 423-427.

[38] Just R, Jalali D, Ernst M D. Defects4J: A database of existing faults to enable controlled testing studies for Java programs[C]// Proceedings of the 2014 International Symposium on Software Testing and Analysis. ACM, 2014: 437-440.

[39] Weimer W, Fry Z P, Forrest S. Leveraging program equivalence for adaptive program repair: Models and first results[C]//2013 28th IEEE/ACM International Conference on Automated Software Engineering (ASE). IEEE, 2013: 356-366.

[40] Kim D, Nam J, Song J, et al. Automatic patch generation learned from human-written patches[C]// International Conference on Software Engineering. 2013.

[41] Monperrus M. A critical review of automatic patch generation learned from human-written patches: essay on the problem statement and the evaluation of automatic software repair[C]//Proceedings of the 36th International Conference on Software Engineering. ACM, 2014: 234-242.

[42] DeMarco F, Xuan J, Le Berre D, et al. Automatic repair of buggy if conditions and missing preconditions with SMT[C]//Proceedings of the 6th International Workshop on Constraints in Software Testing, Verification, and Analysis. ACM, 2014: 30-39.

[43] Cadar C, Dunbar D, Engler D R. KLEE: Unassisted and Automatic Generation of High-Coverage Tests for Complex Systems Programs[C]//OSDI. 2008, 8: 209-224.

[44] Wang S, Wen M, Mao X, et al. Attention Please: Consider Mockito when Evaluating Newly Released Automated Program Repair Techniques[C]//Proceedings of the 23rd International Conference on Evaluation and Assessment in Software Engineering (EASE). ACM, 2019.

[45] Martinez M, Durieux T, Sommerard R, et al. Automatic repair of real bugs in java: A large-scale experiment on the defects4j dataset[J]. Empirical Software Engineering, 2017, 22(4): 1936-1964.

[46] Durieux T, Cornu B, Seinturier L, et al. Dynamic patch generation for null pointer exceptions using metaprogramming[C]//Software Analysis, Evolution and Reengineering (SANER), 2017 IEEE 24th International Conference on. IEEE, 2017: 349-358.

[47] Matias Martinez and Martin Monperrus. 2018. Ultra-Large Repair Search Space with Automatically Mined Templates: the Cardumen Mode of Astor. In Proceedings of the 10th International Symposium on Search-Based Software Engineering (SSBSE '18). Cham, 65–86.

[48] Sobreira V, Durieux T, Madeiral F, et al. Dissection of a bug dataset: Anatomy of 395 patches from Defects4J[C]//2018 IEEE 25th International Conference on Software Analysis, Evolution and Reengineering (SANER). IEEE, 2018: 130-140.

[49] Wen M, Chen J, Wu R, et al. An empirical analysis of the influence of fault space on search-based automated program repair[J]. arXiv preprint arXiv:1707.05172, 2017.

[50] Ye H, Martinez M, Monperrus M. A comprehensive study of automatic program repair on the QuixBugs benchmark[J]. arXiv preprint arXiv:1805.03454, 2018.

[51] Koyuncu A, Bissyandé T F, Kim D, et al. Impact of tool support in patch construction[C]//Proceedings of the 26th ACM SIGSOFT International Sym-posium on Software Testing and Analysis. ACM, 2017: 237-248.

[52] Motwani M, Sankaranarayanan S, Just R, et al. Do automated program repair techniques repair hard and important bugs?[J]. Empirical Software Engineer-ing, 2018, 23(5): 2901-2947.

[53] Wang Y, Meng N, and Zhong H. An Empirical Study of Multi-Entity Changes in Real Bug Fixes[C]// In: Proceedings of IEEE International Conference on Software Maintenance and Evolution. IEEE, 2018: 316-327.

[54] Inozemtseva L, Holmes R. Coverage is not strongly correlated with test suite effectiveness[C]//Proceedings of the 36th International Conference on Software Engineering. ACM, 2014: 435-445.

[55] Gligoric M, Groce A, Zhang C, et al. Guidelines for coverage-based comparisons of non-adequate test suites[J]. ACM Transactions on Software Engineering and Methodology (TOSEM), 2015, 24(4): 22.

[56] Just R, Jalali D, Inozemtseva L, et al. Are mutants a valid substitute for real faults in software testing?[C]//Proceedings of the 22nd ACM SIGSOFT International Symposium on Foundations of Software Engineering. ACM, 2014: 654-665.

[57] Chekam T T, Papadakis M, Le Traon Y, et al. An empirical study on mutation, statement and branch coverage fault revelation that avoids the unreliable clean program assumption[C]//2017 IEEE/ACM 39th International Conference on Software Engineering (ICSE). IEEE, 2017: 597-608.

[58] Madeiral F, Urli S, Maia M, et al. Bears: An Extensible Java Bug Benchmark for Automatic Program Repair Studies [C]//2019 IEEE 26th International Conference on Software Analysis, Evolution and Reengineering (SANER). IEEE, 2019.

[59] Saha R, Lyu Y, Lam W, et al. Bugs. jar: a large-scale, diverse dataset of real-world java bugs[C]//2018 IEEE/ACM 15th International Conference on Mining Software Repositories (MSR). IEEE, 2018: 10-13.

[60] Yu Z, Martinez M, Danglot B, et al. Alleviating patch overfitting with automatic test generation: a study of feasibility and effectiveness for the Nopol repair system[J]. Empirical Software Engineering, 2019, 24(1): 33-67.

[61] Tu F, Zhu J, Zheng Q, et al. Be careful of when: an empirical study on time-related misuse of issue tracking data[C]//Proceedings of the 2018 26th ACM Joint Meeting on European Software Engineering Conference and Symposium on the Foundations of Software Engineering. ACM, 2018: 307-318.

[62] Rodríguez-Pérez G, Robles G, González-Barahona J M. Reproducibility and Credibility in Empirical Software Engineering: A Case Study based on a Systematic Literature Review of the use of the SZZ algorithm[J]. Information and Software Technology, 2018.